\documentstyle[aps,epsfig]{revtex}
\newcommand{\be}{\begin{equation}}
\newcommand{\ee}{\end{equation}}
\newcommand{\bea}{\begin{eqnarray}}
\newcommand{\eea}{\end{eqnarray}}

\begin{document}
\tightenlines
\title{Parton Production Via Vacuum Polarization}
\author{Dennis~D.~Dietrich$^a$, Gouranga~C.~Nayak$^b$, and Walter~Greiner$^a$}
\address
{$^a$\small\it{Institut f\"ur Theoretische Physik,
J. W. Goethe-Universit\"at,
60054 Frankfurt am Main, Germany},\\
 $^b$\small\it{T-8,
Theoretical Division, Los Alamos National Laboratory, Los Alamos, NM
87545, USA}}
\maketitle
\begin{abstract}
We discuss the production mechanism of partons via vacuum polarization during the very early, gluon dominated phase of an ultrarelativistic heavy-ion collision in the framework of the background field method of quantum chromodynamics.
\end{abstract}
\bigskip
For high matter and/or energy densities a deconfined state of matter, the 
so-called quark-gluon plasma (QGP) has been predicted to exist \cite{qgp}. 
Its production is expected for ultrarelativistic heavy-ion collisions 
(URHIC) at RHIC and LHC. No direct signatures for its presence are 
experimentally accessible. In order to be able to decide which indirect ones
could be available, detailed knowledge about the production and evolution of 
the (QGP) is needed.

At the beginning, an URHIC is expected to be dominated by gluonic degrees of 
freedom. The 
gluons are likely to be so abundant that they can be described as a classical 
vector potential $A$. 
We are now interested in finding out whether the production of 
quark-antiquark pairs (see Fig.(1)) or the production of gluon pairs (see 
Fig.(2)) from the field dominates. The concise way 
to handle a classical vector potential and quantum fluctuations 
simultaneously is the background field method \cite{bfm} of quantum 
chromodynamics (QCD). In it, the gauge field $B$ is split into a classical 
vector potential $A$ plus quantum fluctuations $Q$: $B\rightarrow A+Q$. The 
classical background field $A$ is equal to the expectation value of the 
entire field $B$: $<B>=A$. The quantum part contains all fluctuations of 
$B$: $[dB]=[dQ]$; its expectation value is equal to zero: $<Q>=0$. The gauge 
transformation properties of the entire gauge field $B$ have got to be 
redistributed, so that the classical vector potential $A$ satisfies the same 
relations under the gauge transformation $U$ as $B$ beforehand:
$
T^aA^a_\mu\rightarrow T^aA^{'a}_\mu
=
UT^aA^a_\mu U^{-1}-\frac{i}{g}(\partial_\mu U)U^{-1}.
$
Only for this choice does the background field $A$ satisfy the classical 
Yang-Mills equations. In order to conserve the sum, the field of the quantum 
fluctuations $Q$ has got to transform according to:
$
T^aQ^a_\mu\rightarrow T^aQ^{'a}_\mu=UT^aQ^a_\mu U^{-1}.
$
This set of gauge transformations is called type-(I)-gauge transformations 
\cite{iz}.
At the end, a gauge fixing term, the so-called background field gauge, is 
chosen which leaves the resulting effective action gauge invariant under 
type-(I)-gauge transformations:
$
\partial^\mu Q^a_\mu+gf^{abc}A^{b\mu}Q^c_\mu=C^a.
$
We still have the choice to treat the production processes 
perturbatively or non-perturbatively. As we finally would like to determine 
the space-time structure of the classical field by means of a self-consistent 
calculation, we choose the perturbative approach (see Figs.(1) and (2)) as analytical non-perturbative results 
exist only for very few special choices of the form of the field. The results 
of the self-consistent calculations will be presented elsewhere.

The range of applicability of perturbative calculations is limited. The 
described particles must have an energy $p^0$ much greater than the typical 
field strength $A$ of the classical vector potential multiplied by the 
coupling constant $g$. As we are going to be dealing with a decaying field 
this is always satisfied if we take $p^0$ above $gA_{in}$, where $A_{in}$ is 
the initial field strength. Further, the product $gA$ has got be much larger 
than the other non-perturbative scale of QCD, $\Lambda_{QCD}$. This results in 
the fact that the approach becomes invalid after a maximum time $t_f$ after 
which the quantity $gA$ reaches the scale $\Lambda_{QCD}$.
If this description of the gluonic sector is sufficient, depends on whether
quantum gluons are important for momenta below $gA$. 
Generally speaking, for a field that decays more rapidly as
compared to another with the same initial field strength, soft quantum
corrections are less important.
 For the 
quarks there is no means of description for the low momentum sector here. 
This feature has got to be added in a later study. 
For field strengths above
$\Lambda_{QCD}$ one would expect that a description via the constant-field 
Schwinger approach \cite{jss} would be adequate, but there the relatively big 
slopes of the decaying field are ignored totally. Nevertheless, a comparison 
to calculations involving that formula would be interesting.

Even for our perturbative approach the general expressions are very 
complicated \cite{PRD}. That is why we choose a special form for a purely 
time-dependent field in order to get some insight into the behavior of the 
obtained source terms for quarks and gluons:
$
A^{a3}(t)=A_{in}e^{-|t|/t_0},~t_0>0,~a=1,...,8,
$
and all other components are equal to zero. 
Many other forms could have been taken. Its time structure is similar to one 
obtained from a different numerical study \cite{num}.
Using this field, one obtains the following expressions for the source terms:

\bea
\frac{dW_{q\bar{q}}}{d^4x d^3k}
=
16\frac{\alpha_S}{(2\pi)^2}
(A_{in})^2
e^{2 i \omega t}
e^{-|t|/t_0}
\frac{t_0}{1+4\omega^2t_0^2}
\frac{m_T^2}{\omega^2},
\eea

and:

\bea
\frac{dW_{gg}}{d^4xd^3k}
=
\frac{24\alpha_S}{(2\pi)^2}(A_{in})^2e^{2ik^0t}e^{-|t|/t_0}
\frac{t_0}{1+4(k^0)^2t_0^2}(-3-\frac{k_T^2}{(k^0)^2})
~\nonumber \\
+
\frac{36\alpha_S^2}{2\pi}(A_{in})^4e^{2ik^0t}e^{-2|t|/t_0}
\frac{t_0}{1+(k^0)^2t_0^2}\frac{1}{(k^0)^2}.
\eea

Note, that the real part of these expressions has got to be taken. 
The contribution of the interference term for the production of gluons
vanishes for all fields of the form $A^{a\mu}(x)=A^{a\mu}_{in}f(x)$.

For LHC one expects the following parameters which are consistent with a predicted energy density of $1TeV/fm^3$: for the initial value of the 
field $A_{in}=1500MeV$, for the coupling constant $\alpha_S=0.15$, and for the 
decay-time $t_0\in\{0.1fm,1.0fm\}$.
It can be checked that for $gA_{in}>>\Lambda_{QCD}$ the production 
of quantum particles after the final time of applicability $t_f$ can be 
neglected \cite{dng}.

In order to decide which parton production channel dominates,
it is recommendable to investigate the ratio of produced gluon pairs over 
produced quark-antiquark pairs in the kinematic region mentioned above
for LHC. For lower momenta gluon production clearly dominates over
quarks production, which is evident from an extra $1/(k^0)^2$ in the second
term in Eq. (2). For the momentum range we are 
interested in ($p_t \simeq $ 2 GeV) all kind of scenarios are possible
depending upon the value of $t_0$. The ratio of gluon to quark source
term as a function of $t_0$ is shown in Fig.(3) for $p_t$ = 2 GeV at LHC.

There quark production prevails over gluon production for smaller values of 
the decay time $t_0$.
For larger values of $t_0$, the gluon channel 
is the dominant one. 
A selfconsistent calculation yielding the decay-time as an output has got to 
decide between the scenarios.

That result would also be important in another sense.
In the case of the dominance of gluon pair creation additional 
fermions and antifermions have got to be produced in subsequent 
inelastic collisions, not to forget about intermediate elastic 
rescatterings \cite{rndg}. These quarks can play a crucial role
in hadron production at RHIC and LHC.
Of course all this would have to be explored in extensive calculations 
where fragmentations of gluons have got to be taken into account.


\newpage
\begin{figure}[h]
\begin{center}
\epsfig{figure=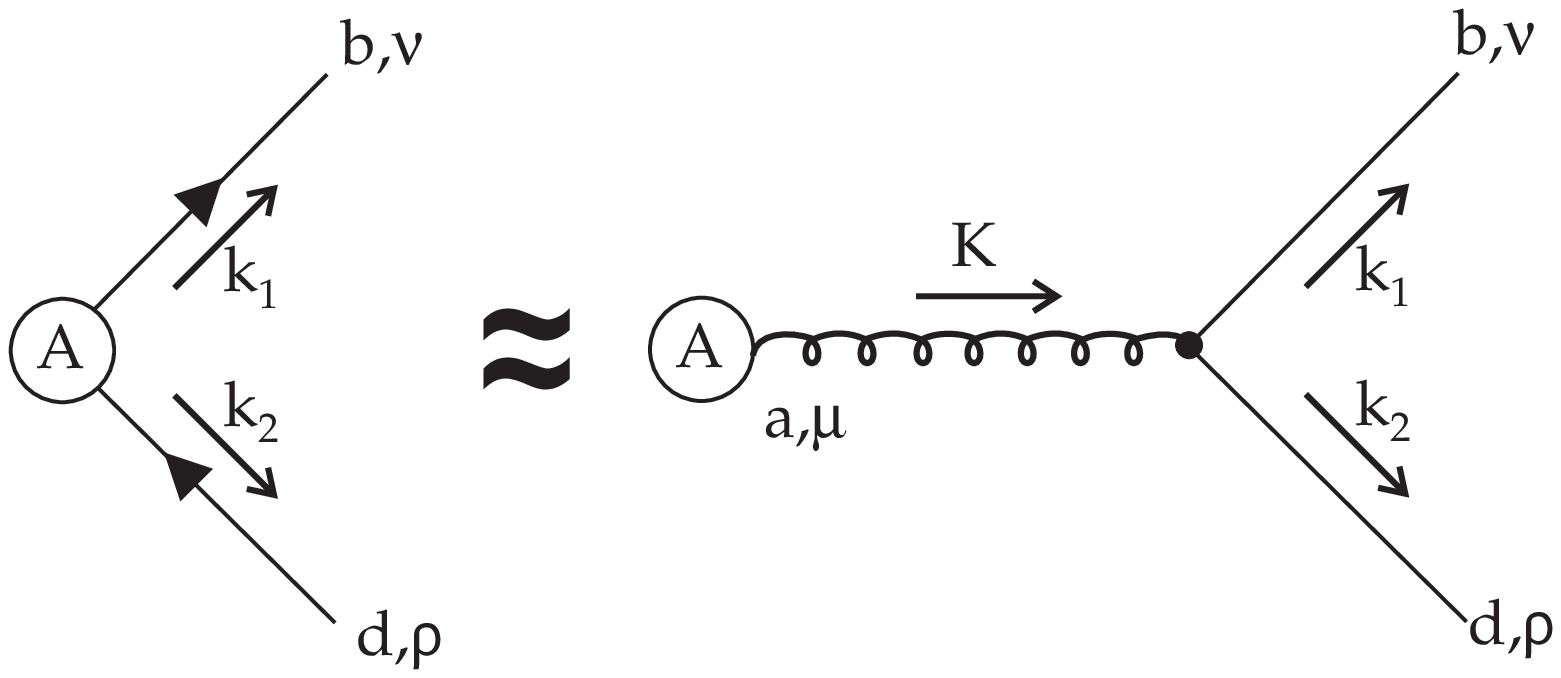, width=5cm} \\
{\it Fig.1: The production of a quark-antiquark pair by coupling to the field once 
is the dominant contribution for particles with an energy higher than the 
value of the field multiplied by the coupling constant.}
\end{center}
\end{figure}

\vspace{1cm}

\begin{figure}[h]
\begin{center}
\epsfig{figure=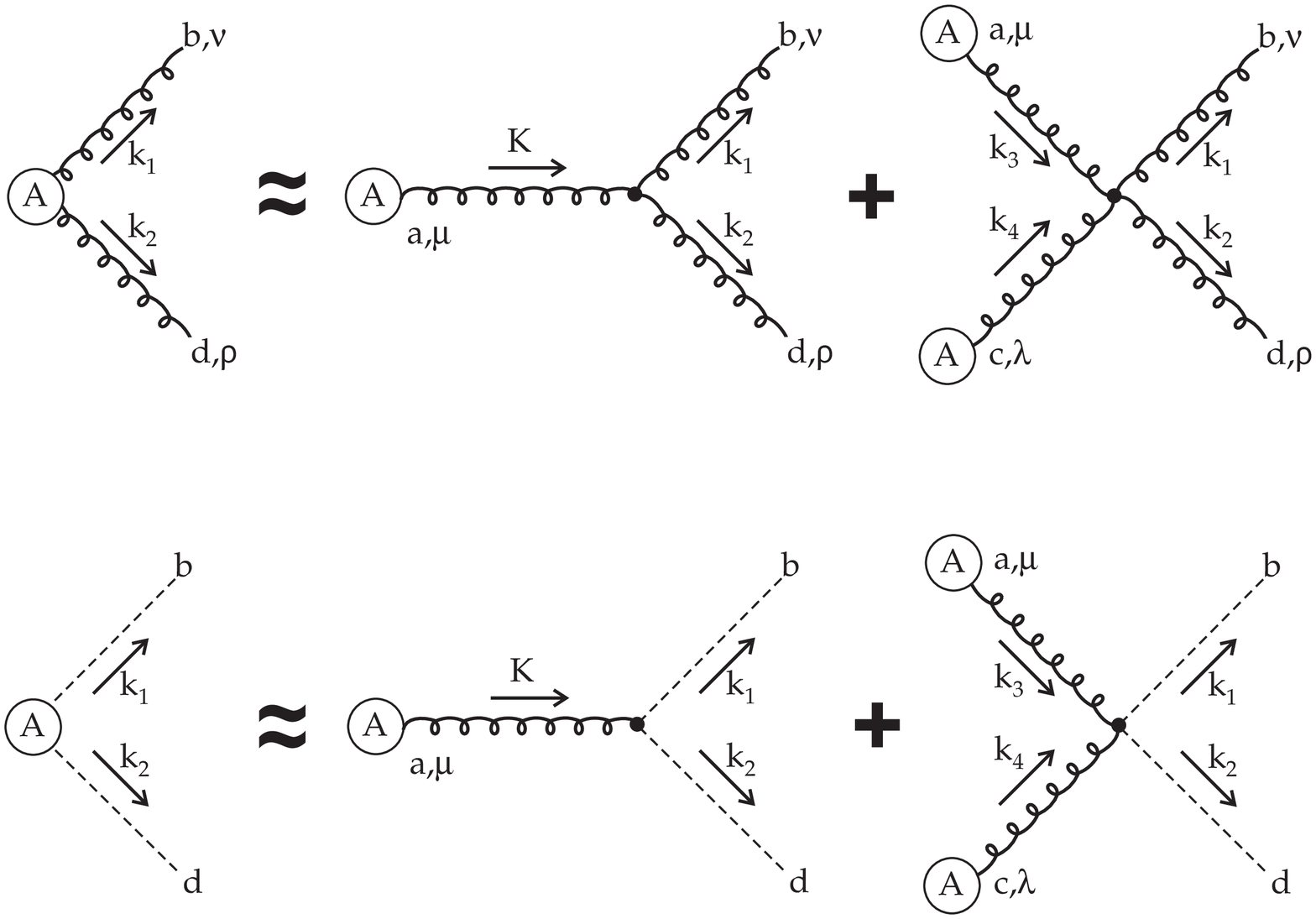, width=7.5cm} \\
{\it 
Fig.2:
The production of a gluon pair by coupling to the field at one vertex is the 
dominant contribution for particles with an energy higher than the value of 
the field multiplied by the coupling constant. The chosen gauge is not 
physical, so the corresponding ghost contributions have got to be taken into 
account.
}
\end{center}
\end{figure}

\vspace{1cm}

\begin{figure}[h]
\begin{center}
\epsfig{figure=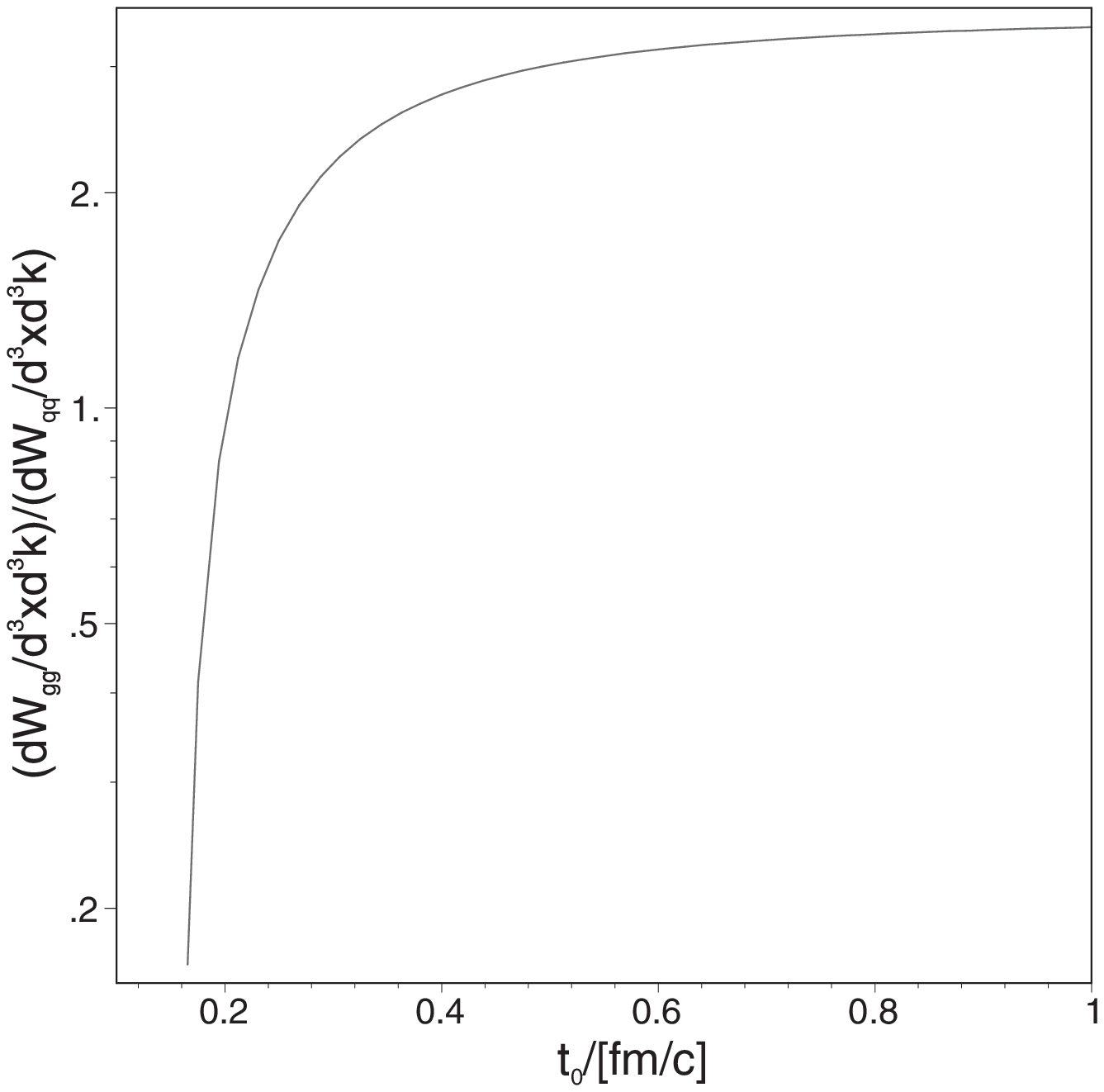, width=7.5cm} \\
{\it 
Fig.3:
Time integrated source term for gluon production divided by the one for quark 
production versus the decay time $t_0$.
}
\end{center}
\end{figure}
\end{document}